\documentclass[authoryear]{elsarticle}

\usepackage[top=1.25in, bottom=1.25in, left=1.5in, right=1.5in]{geometry}
\usepackage{lineno}
\usepackage{soul}
\usepackage{color}
\usepackage{tikz}
\usetikzlibrary{quotes,angles}
\usepackage[pagebackref=true,
            colorlinks=true,
            bookmarks=true,
           ]{hyperref}   
           
\usepackage[most]{tcolorbox}           
\usepackage{graphicx}
\usepackage{epstopdf}
\DeclareGraphicsExtensions{.eps}
\usepackage{siunitx}

\usepackage{empheq}
\usepackage{amssymb}
\usepackage{array}
\usepackage{amsmath, scalerel}
\usepackage{leftidx}

\usepackage{amssymb}
\usepackage{gensymb}

\usepackage{amsthm}

\usepackage{mathptmx}

\usepackage{amsfonts}
\usepackage{multicol}
\usepackage{mathrsfs}
\usepackage{tensor}
\usepackage{subfig}
\usepackage{hhline}
\usepackage{upgreek}
\usepackage{cancel}
\usepackage{ulem}
\newcommand*\circled[1]{\tikz[baseline=(char.base)]{
            \node[shape=circle,draw,inner sep=1pt] (char) {#1};}}

\usepackage{multirow}
\usepackage{setspace}

\usepackage[flushleft]{threeparttable}
\usepackage{makecell,booktabs}

\usepackage{amsmath,scalerel}

\newcommand{\tightoverset}[2]{\mathop{#2}\limits^{\vbox to -.5ex{\kern-0.75ex\hbox{$#1$}\vss}}}
\modulolinenumbers[5]

\journal{Elsevier}

\begin{document}

\allowdisplaybreaks[4]

\begin{frontmatter}
\title{Influence of orientation-dependent lath martensite yielding on the hardening behavior of quenched martensitic steels}

\author[1,2]{V. Rezazadeh}
\author[1]{\corref{cor}R.H.J. Peerlings}\ead{r.h.j.peerlings@tue.nl}
\author[1]{J.P.M. Hoefnagels}
\author[3]{F. Maresca}
\author[1]{M.G.D. Geers}

\address[1]{Department of Mechanical Engineering, Eindhoven University of Technology (TU/e), P.O.Box 513, 5600 MB Eindhoven, The Netherlands}
\address[2]{Materials Innovation Institute (M2i), P.O.Box 5008, 2600 GA Delft, The Netherlands}
\address[3]{Faculty of Science and Engineering, P.O.Box 72, 9747 AG  Groningen, The Netherlands}

\cortext[cor]{Corresponding author.}

\begin{abstract}
The onset of plasticity in quenched martensitic microstructures is characterized by a low initial yield stress followed by an extremely strong initial hardening response, and then a sudden hardening saturation. Literature attributes this behavior to residual stresses and dislocations inherited from the martensitic transformation, or to microstructural heterogeneities causing strength differences among the grains. Here, we argue that orientation-dependent yielding of lath martensite due to inter-lath sliding, which induces a substructure boundary sliding mechanism that may also contribute significantly to the observed behavior. To demonstrate this, we systematically study its quantitative contribution to the elasto-plastic transition behavior in a numerical microstructural model. In the simulations, we employ an effective laminate model for the martensite packets which takes into account the yielding anisotropy due to sliding along the packet's habit plane orientation. To account for the effect of carbon content, martensitic microstructures with different levels of lath strength are considered. It is shown that the martensite packets with a favorable habit plane orientation start to yield earlier compared to those with an unfavorable orientation, which initially remain elastic. As a consequence, the macro-scale response of the microstructures exhibits a low yield stress, followed by a significant degree of initial hardening which continues until the saturation stress level is approached. The apparent work hardening rate depends on the contrast between the in-habit plane and out-of-habit plane yield strength used in the model. Moreover, our simulations are able to qualitatively capture other observations reported in the literature regarding the high initial hardening behavior of quenched martensitic steels, e.g. that the initial plastic yielding is independent of carbon content and thus of lath strength, and that the uniform elongation increases by increasing the macroscopic strength.

\end{abstract}

\end{frontmatter}


\section{Introduction}
The use of lath martensite as one of the main constituents in virtually all advanced high strength steel grades has triggered great interest in a better understanding of the physical mechanisms that control its strength and ductility.
A well-established characteristic  of as-quenched martensitic steels is their high initial work hardening rate. Notably, the magnitude of the initial strain hardening exceeds the limit for the dislocation storage mechanism ($ \sim E/50$) reported for polycrystalline materials \citep{KOCKS2003171}. 
Local plasticity, or so-called 'micro-yielding', of isolated grains at low stress followed by the sequential yielding of additional grains have been suggested as the mechanisms underlying the continuous, extended elasto-plastic transition \citep{zaccone1993elastic}. This behavior is further characterized by the following observations reported in literature:

\begin{itemize}
\item The initial yield strength is independent of the carbon content. It has been suggested that the onset of plastic instability is triggered by the presence of soft microstructural elements, the yield strength of which is nearly independent of the chemical composition \citep{takenouchi2022relation,allain2012toward, nakashima2007yielding,zhou2020mechanical}.  
\item The magnitude of the high initial hardening, and the Bauschinger effect, is increased with increasing carbon content. Similarly, the uniform elongation increases by increasing the strength level (and hence by carbon content), contrary to the more common trade-off between strength and ductility (i.e. the 'banana curve' showing strength versus ductility in steels) \citep{krauss2001deformation, hutchinson2018plastic, KRAUSS199940, allain2012toward, wang2021transitions}.
\item The continuous yielding behavior is sustained up to the point of necking \citep{allain2012toward, hutchinson2015yielding}. In other words, the anisotropic yielding of microstructural constituents occurs at high strains, close to the necking strain. 
\item In case of intermediate to high tempering of martensite, the high initial work hardening rate is diminished \citep{krauss2017tempering, MALHEIROS201738, wang2020strain,zhou2020mechanical}. 
\end{itemize}
\begin{figure}[ht!]
\centering
  \includegraphics[width=0.6\linewidth]{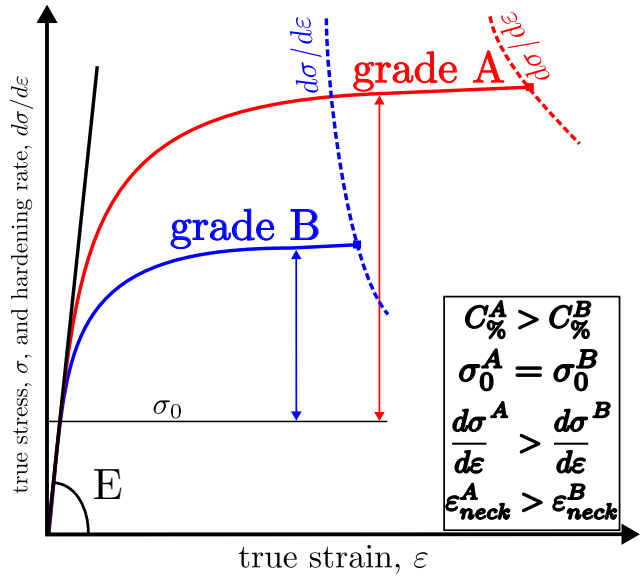}
  \caption{Sketch of the typical hardening behavior of martensite, with a low, composition-independent yield stress, $\sigma_{0}$, followed by a very strong initial hardening which levels off towards saturation. The hardening response does depend on carbon content as shown by the two grades sketched, where grade \textit{A} has a higher carbon content than grade \textit{B}. The dashed curves represent the hardening rate, $d \sigma / d \varepsilon $, of the two grades. Their intersection with their corresponding (solid) $\sigma-\varepsilon$ curve marks the point of necking as predicted by Consid\`{e}re's criterion. For the stronger grade \textit{A} necking occurs later than for the softer (low C) grade \textit{B}.}
  \label{fig:lit}
\end{figure} 

	The sketch in Figure \ref{fig:lit} of the elasto-plastic response of two grades, \textit{A} and \textit{B}, with different carbon content summarizes most of these aspects. Also indicated in the sketch are the work hardening rate, $d \sigma / d \varepsilon $, and the point of necking, which according to Consid\`{e}re's criterion is given by the intersection of the $\sigma- \varepsilon$ and $d\sigma /d\varepsilon - \varepsilon$ curves, for both grades.  

	An investigation of the available literature reveals that there is no single explanation that addresses all the above mentioned characteristics of lath martensite behavior. The possible mechanisms underlying this behavior, as pointed out in the literature, can be grouped into two main categories. According to the first category, an essential role is played by the short range (type II) residual internal stresses and the large amount of mobile dislocations inherited from the martensitic transformation \citep{hutchinson2015yielding, hutchinson2018plastic,jo2017plasticity,takenouchi2022relation,Suiker_2007,Suiker_2007b}. The second category highlights the heterogeneities in the microstructure causing intrinsic local variability in strength, such as a non-uniform carbon distribution, variations in the lath sizes or in the internal structure of laths, e.g. in terms of carbide content or dislocation densities \citep{nakashima2007yielding,GHASSEMIARMAKI20133640, MORSDORF2015366, Badinier2015}. 

	An aspect of the lath martensite plasticity which has not received much attention yet in this context is the lath boundary orientation-dependent yielding \citep{kwak2016anisotropy, MICHIUCHI20095283,GHASSEMIARMAKI20133640}. Multiple studies have shown highly localized plasticity occurring at low stress levels at all types of substructure boundaries within the packets, i.e. lath, sub-block, and block boundaries, showing severe sliding of these boundaries  \citep{du2016plasticity,du2019lath,LIU2021116533}. This so-called 'substructure boundary sliding' phenomenon is hypothesized to be triggered either by the transformation or the plastic flow of thin retained austenite films between the martensite laths \citep{Maresca2014,MARESCA2017302,MARESCA2018463,DU2018411, COTAARAUJO2021106445, rezazadeh2022extensive}. Others attributed it to the easy glide of intra-lath dislocations close to and parallel to the lath boundaries \citep{inoue2019slip, morsdorf2016multiple}. 
Regardless of its underlying mechanism, the result is that packets in which the the lath boundaries experience a larger resolved shear stress exhibit a significant amount of plastic strain, whereas packets with less favorably oriented boundaries exhibit the strong and brittle response traditionally associated with martensite. We hypothesize that the apparent strength variation due to this orientation dependence may be the cause of the unique overall response described above, by early yielding of favorably oriented packets, followed by a gradual engagement of the less favorable ones, until the intrinsic lath strength is reached and hence all packets are activated. Indeed, this mechanism can qualitatively and perhaps quantitatively explain most of the reported specific observations related to strong initial hardening behavior.

	The above reasoning motivated us to study the effect of the microscopic, orientation-dependent yielding of martensite packets on the macroscopic initial hardening behavior of martensitic steel. The main scientific question to be addressed by our study is whether the effect is strong enough to be considered as a realistic mechanism underlying the peculiar low initial yield stress followed by extreme initial hardening observed in experiments. To answer this question, we adopt a microstructural modelling approach. Representative volume elements are constructed based on synthetic microstructures in which the individual packets are resolved. The habit plane orientations within the packets are sampled from a uniform distribution. Inside the packets, we employ the effective laminate model proposed by \citet{VRezazadeh2022}. This model assumes the anisotropic yielding of the packets to be due to substructure boundary sliding mechanism, which is modeled via a homogenized laminate of hard and thick layers of lath martensite with embedded thin and soft discrete planes of austenite. In our work, we do not address the narrowing of X-ray diffraction patterns observed in quenched martensitic steels, as it indeed can be due to either the relaxation of type II residual stresses (due to strains induced by the martensitic phase transformation), or the change in the initial dislocation density which is inherited from the martensitic transformation \citep{ungar2017composite,hutchinson2018plastic,odor2020deformation, wang2021transitions}. Moreover, the observed strong Bauschinger effect is not studied here, since there is a lack of experimental information on the substructure boundary sliding phenomenon under non-monotonic loading conditions.

\begin{figure}[ht!]
\centering
  \includegraphics[width=1\linewidth]{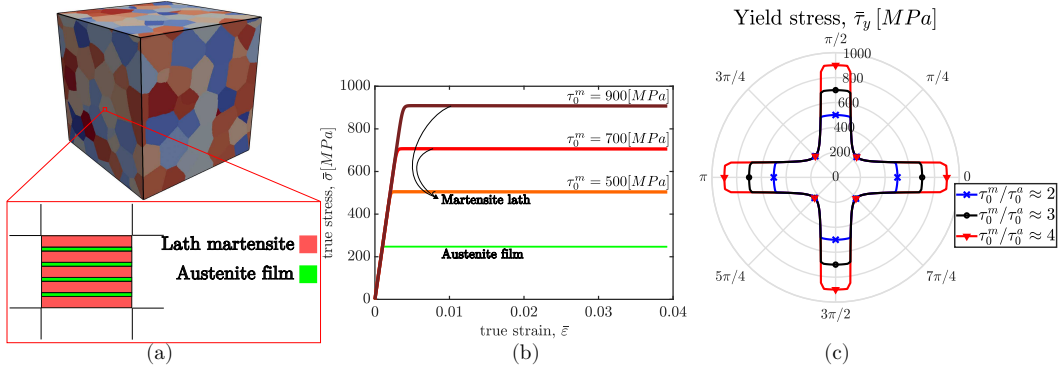}
  \caption{a) A three-dimensional RVE of a virtual martensitic steel containing $180$ packets with a random habit plane orientation. The shown RVE is taken from an ensemble of $10$ statistically equivalent RVEs. An effective laminate model proposed by \citep{VRezazadeh2022} is used in each material point, with a random habit plane orientation per packet. b) Input responses of the austenite films and lath martensite in the three cases considered: a reference case and harder and softer laths; the austenite films have the same properties in all three cases. c) The yield stress of the effective laminate model obtained for three values of the yield stress ratio $\tau^{m}_{0}/\tau^{f}_{0}$, where $\tau^{m}_{0}$ is the yield stress of martensite lath, and  $\tau^{a}_{0}$ is the yield stress of austenite. }
  \label{fig:micro}
\end{figure} 
\section{Microstructural model}
A three-dimensional representative volume element (RVE) representing a virtual single-phase martensitic microstructure is employed in our study, as shown in Figure \ref{fig:micro}a. It comprises $180$ packets which were generated by a Voronoi tessellation based on a random set of seed points. The microstructure shown in Figure \ref{fig:micro}a is one realization taken from an ensemble of 10 RVEs used here. To account for the boundary sliding behavior of lath martensite, a homogenized laminate model proposed by \citet{VRezazadeh2022} has been employed. 
In the model, every material point is endowed with the effective response of an infinite periodic laminate representing martensite laths and soft films of retained austenite -- see Figure \ref{fig:micro}a. Plastic yielding of the laths is governed by isotropic, von Mises, plasticity, whereas a dedicated planar plasticity model -- isotropic in the plane of inter-lath austenite film -- is employed for the austenite films. The plastic response of the material point is determined by the competition between these two plastic mechanisms. This constitutive model only requires the habit plane orientation of the packet and the plastic properties of the two phases. Hence, crystallographic effects or texture are not taken into account, consistent with statements made in the literature that it is too weak to be significant for the overall initial yielding response \citep{KOCKS2003171,allain2012toward}.

In the microstructure shown in Figure \ref{fig:micro}a, a random habit plane orientation is assigned to each martensite packet, according to a uniform orientation distribution. 
Although the two phases in the model are isotropic, a significant degree of plastic anisotropy is inherited from the orientation of the retained austenite films, combined with the contrast in properties of the phases. 
This is shown in the polar plot in Figure \ref{fig:micro}c, where the yield limit of the model as a function of the angle between the normal of the films and the applied tensile load, $\theta$, is given. The yield criterion used here is defined based on the criterion employed in \citet{VRezazadeh2022}.
The yield strength of the retained austenite film is constant, while three different levels of strength are considered for the laths -- a reference case and two cases with comparatively softer and harder laths, see Figure \ref{fig:micro}b. Due to the segregation of C to the lath boundaries, an increase in carbon content only makes the laths harder without changing the properties of the retained austenite \citep{hutchinson2011microstructures,Zhang2012109,morsdorf2021carbon}. 
Yet, it has been reported that the volume fraction of retained austenite increases with an increase in carbon content \citep{saeglitz1997deformation, morito2011carbon}. 
This has been taken into account in the model by decreasing the distances between the austenite films accordingly \citep{VRezazadeh2022}. The corresponding values are $d/l= 0.05/0.08/0.1$, for the simulation with soft, reference, and hard laths, respectively. 

As shown in Figure \ref{fig:micro}b, an ideally plastic response is considered for both phases, i.e. no hardening included \citep{delince2007structure, MICHIUCHI20095283}. This is done on purpose, as our hypothesis is that the overall response of the material, i.e.  of the RVE, will nevertheless exhibit hardening due to the interactions among of the packets with differently orientated habit planes/austenite films. The assumption of ideal plasticity for both phases enables us to systematically track the origin of any observed hardening induced in the simulations. The simulations were done using a FFT-based spectral solver implemented in the DAMASK simulation package \citep{ROTERS2019420}. An average horizontal tensile load is applied incrementally to the RVE. A $256 \times 256 \times 256$ point spatial discretization grid was employed in all simulations.

\section{Response of martensitic microstructures with different lath strengths}
\begin{figure}[ht!]
\centering
  \includegraphics[width=0.7\linewidth]{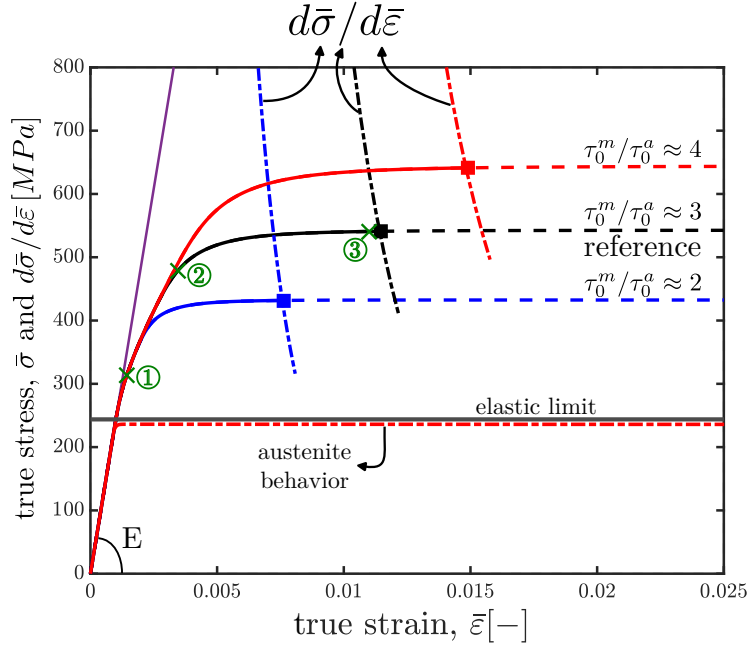}
  \caption{ The macroscopic equivalent true stress-strain response obtained from the simulations performed with different lath strengths. The yield limit is determined based on a $2\%$ drop with respect to the elastic slope. The dashed curves show the work hardening rate, $d\sigma/d\varepsilon$, which has been computed to determine the necking point for each stress-strain curve. The necking strains are marked with square markers. Three strain steps, tagged \textcircled{1} -- \textcircled{3}, are sampled from the response of the reference case, $\tau_{0}^{m}/\tau_{0}^{a}\approx 3$, discussed further in the text.  }
  \label{fig:GlobRes}
\end{figure}

In this section, the results are presented. 
Figure \ref{fig:GlobRes} shows the macroscopic true (logarithmic) stress-strain response of the three simulation cases done with the microstructural model shown in Figure \ref{fig:micro}. The work hardening rate, $d\bar{\sigma}/ d\bar{\varepsilon}$, has been computed and also plotted for each case to estimate the point of necking by Consid\`{e}re's criterion. The effective elastic limit, i.e. initial yield point, is arbitrarily defined as the point at which the slope of the stress-strain curves has decreased $2\%$ relative to the elastic stiffness, i.e. $d\bar{\sigma}/ d\bar{\varepsilon} = 0.98E$. 
It is observed that the initial yield point of all three cases coincides and its magnitude corresponds to the initial yield strength of the boundary sliding mechanism, $\tau^{a}_{0}\approx 250$ MPa, see the red dash-dotted curve in Figure \ref{fig:GlobRes}b.
Immediately following initial yield, all cases show an extremely high initial work hardening rate - on the order of magnitude of the elastic modulus. 
It is noticeable that, whereas the hardening curves observed for the three lath strengths initially coincide, they start to deviate at approximately $0.2\%$ strain.
Beyond this point, a higher work hardening rate is observed for a larger value of lath strength.
Also note that the uniform elongation increases with the lath strength - and hence with the strength of the martensitic microstructures.
All the above observations are, qualitatively, following the experimental observations reported in the literature on martensitic steels \citep{KRAUSS199940, allain2012toward,hutchinson2011microstructures, hutchinson2015yielding,jo2017plasticity}.

\begin{figure}[ht!]
\centering
  \includegraphics[width=1\linewidth]{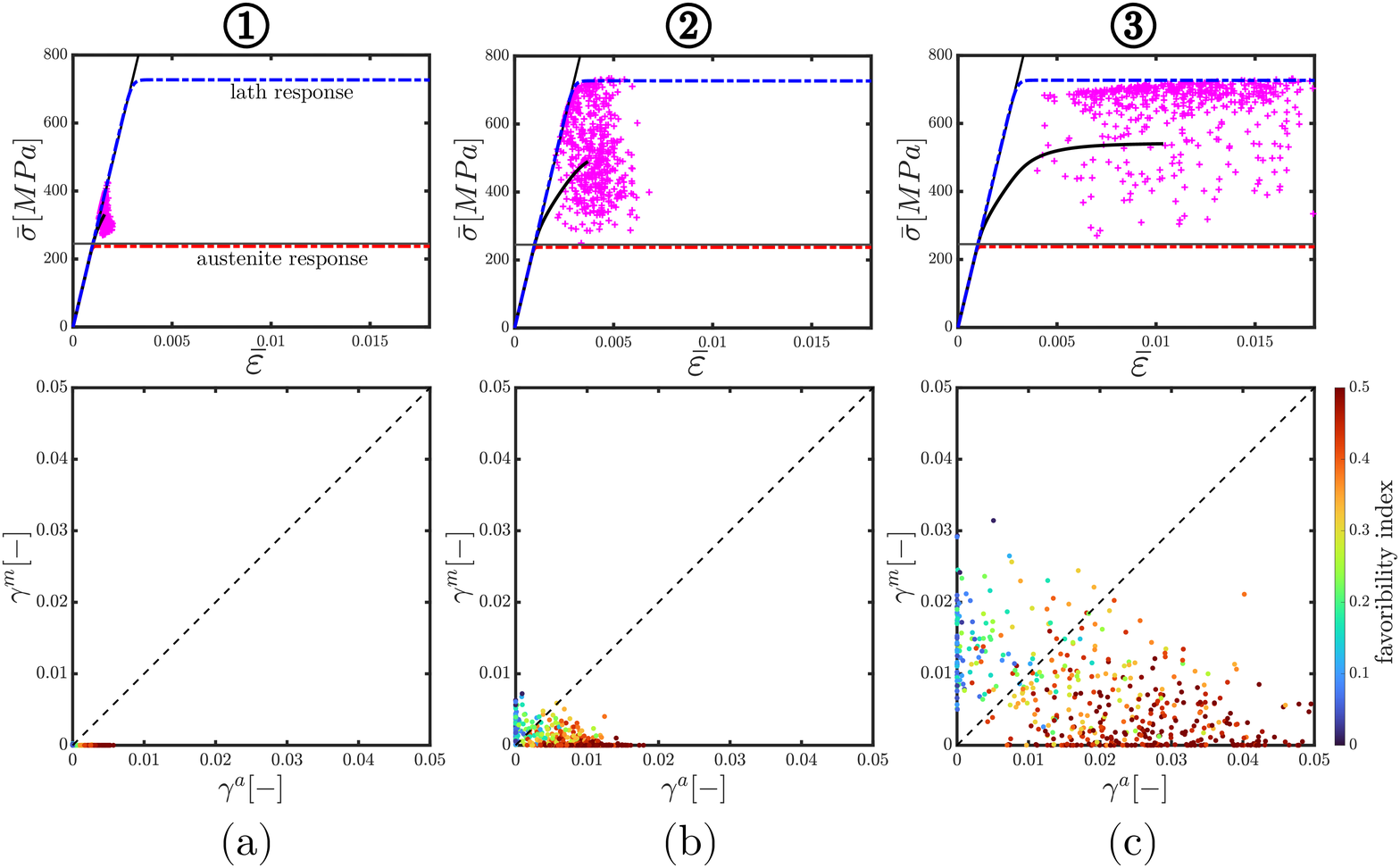}
  \caption{top row: equivalent mean true stress versus strain for each individual packet (magenta data points) of the reference microstructure, the presented RVE in Figure \ref{fig:micro}, at the values of the applied strain tagged \textcircled{1} -- \textcircled{3} in Figure \ref{fig:GlobRes}. The macroscopic average response (see Figure \ref{fig:GlobRes}), the austenite and lath martensite input response, and the initial elastic slope are also plotted. Bottom row: plastic activity of the martensite laths, $\gamma^{m}$, vs. that of the inter-lath retained austenite film, $\gamma^{a}$, for each packet. The color indicates the favorability index ($FI$) for substructure boundary sliding. }
  \label{fig:Dist}
\end{figure} 
To further rationalize the simulation results, the stress-strain behavior of the individual packets in the reference case, $\tau_{0}^{m}/\tau_{0}^{a} \approx 3$, of the RVE (Figure \ref{fig:micro}a) for the three applied strains marked in Figure \ref{fig:GlobRes} is shown in Figure \ref{fig:Dist}-top row. 
Point \circled{1} is just beyond the onset of plasticity, in the regime with very high hardening rate. Point \circled{2} is in the transition to a milder hardening, and \circled{3} is taken close to saturation. 
For comparison, the input response of austenite (governing boundary sliding) and lath martensite are plotted by red and blue dashed lines, respectively.  
The bottom plots in Figure \ref{fig:Dist} compare the type of plastic activity in each packet, i.e. whether it is due to sliding of the austenite films, $\gamma^{a}$, or due to plasticity in the martensite lath, $\gamma^{m}$. The color indicates the favorability index ($FI$), which describes the maximum intensity of the shear stress acting on the austenite films in any in-plane shearing direction caused by the applied normal stress: it is computed as,
\begin{equation}
FI= \mathrm{cos}(\theta)\cdot \mathrm{sin}(\theta),
\end{equation}
in which $\theta$ is the angle between the habit plane normal and the direction of the applied tensile stress. The favorability index ($FI$) is different from the conventional Schmid factor, as $FI$ projects the shear stress on its acting direction on the plane and not on the particular slip system direction. A higher $FI$ in a packet, e.g. close to the maximum value of $0.5$, implies a higher likelihood of substructure boundary sliding mechanism being activated.

\begin{figure}[ht!]
\centering
  \includegraphics[width=1\linewidth]{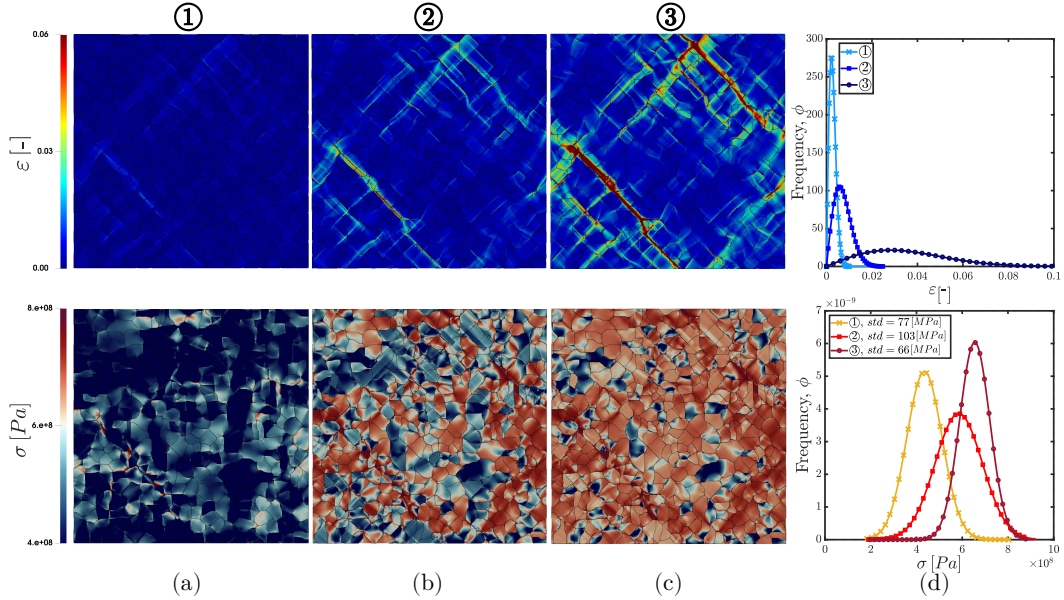}
  \caption{a-c) The equivalent true strain and stress in a microstructure taken from the ensemble at the same applied strains as above. d) The frequency distribution of the same fields obtained from the microstructures shown in a-c. }
  \label{fig:map}
\end{figure}

 Figure \ref{fig:map}a-c shows the equivalent true stress and strain fields in a microstructure taken from the ensemble at the same applied strains as in Figure \ref{fig:Dist}. The frequency distributions of the same fields, are given in \ref{fig:map}d, with the computed standard deviation for the stress distributions. It is observed that the heterogeneity of the strain distribution becomes larger as the applied deformation increases. However, the highest variability in stress is observed in the transition regime, i.e. in point \circled{2}.  

\section{Discussion on the effect of lath martensite anisotropic yielding behavior}
In this section, the results of the simulations presented in the previous section are discussed in detail.
At the early stages of the deformation, point \circled{1} shown in Figure \ref{fig:Dist}a-top, most of the martensite packets are still elastic whereas some with a favorable habit plane orientation, i.e. $IF\approx 0.5$, start to be plastically active.  At this stage, the micro-yielding is only due to sliding of austenite films at lath boundaries, see Figure \ref{fig:Dist}a-bottom in which no activity inside the martensite is shown. As a consequence, the macroscopic plastic yielding initiates at a low stress level necessary to activate the plasticity in the austenite films, i.e. $\tau^{a}_{0}\approx 250$ MPa. The subsequent steep stress-strain curve is the result of the average response of elastic packets and a limited number of plastic ones undergoing sliding. Figure \ref{fig:map}a shows a high variability of stress in the microstructure--standard deviation $std=77$ MPa, even though the strains are low with peak stresses occurring in grains neighboring the first plasticity deforming grains (in which stresses are low). The onset of plasticity at a low stress due to the presence of soft retained austenite has been initially suggested by \citet{zaccone1993elastic}, and supported by the crystal plasticity simulations of \citet{Maresca2014,maresca2016reduced}. The independence of the micro-yielding to carbon content can perhaps be concluded from the experimental observations of \citet{inoue2019slip}, in which, at a similarly low applied tensile strain level (probably at low stress levels), both ultra low and low carbon martensitic microstructures develop slip bands parallel to the lath boundaries oriented at $45^\circ$.

In the transition regime, point \circled{2}, shown in Figure \ref{fig:Dist}b-top, not only favorably oriented packets are plastic, through boundary sliding, but also laths with an unfavorable habit plane orientation start to become plastically active, through plasticity in the laths. This is shown in Figure \ref{fig:Dist}b where several packets with a low $IF$ are active; these are located above the dashed line (with slope $1$) which indicates a higher plastic activity in the lath, $\gamma^{m}$, than in the austenite film, $\gamma^{a}$. Despite the plastic activity of several unfavorably oriented martensite packets, there are also many martensite packets which are still elastic, see Figure \ref{fig:GlobRes}b-top. This results in the apparent high initial effective hardening induced by the microstructure that extends over a large stress range, as observed in Figure \ref{fig:GlobRes}. In this regime, the highest stress variability occurs in the microstructure, $std=103$ MPa, see Figure \ref{fig:map}b\& d. Essentially, the extended elasto-plastic transition is a reflection of the gradual yielding of martensite grains. If the strength contrast between the active and inactive sliding is increased, e.g. by making the lath harder, the magnitude of the work hardening rate increases, which in turn leads to delayed necking and a higher uniform elongation, see Figure \ref{fig:GlobRes}. This is in contrast to the commonly observed trade-off between the strength and ductility in other advanced high strength steels (and other metals). 

As the applied deformation increases, towards point \circled{3} shown in Figure \ref{fig:Dist}c-top, all the packets become plastically active, i.e. no persisting elastic packets. This leads to a saturation of the stress-strain curve as shown in Figure \ref{fig:Dist}c. At this stage, the strain distribution in the microstructure is highly heterogeneous, while the variability in stress is the lowest, $std=66$ MPa, see Figure \ref{fig:map}d. Note that this nearly complete saturation of the hardening behavior occurs already at a very small strains of around $0.5\%$ which is a unique feature for martensite, which is naturally explained by the transition to the regime where all grains become active, whether by substructure boundary sliding or lath plasticity. The strength of the microstructure at this stage is determined by the collective effect of austenite sliding and lath plasticity. This contradicts what has been suggested in literature that the ultimate strength reached in martensitic steels corresponds to the intrinsic strength of the laths, as is clearly visible in Figure \ref{fig:Dist}c where the black line is well below the blue dashed curve.

Based on our finding, it seems that the role of the anisotropic yielding of lath martensite in the high initial hardening behavior is underestimated in the literature. As a counter argument, though, it has been argued that the austenite transforms at the early stages of deformation, and subsequently no longer contributes to plasticity. However, several experimental studies have reported that micro-tensile specimen deformed only by sliding along lath boundaries until failure by complete detachment of the adjacent laths, which corresponds to an extremely high local shear strain of the lath boundary \citep{du2016plasticity,DU2019107646, morito2007comparison}. Moreover, also for experiments on bulk specimens in the experiments of \citet{morsdorf2016multiple,inoue2019slip} strain localization along lath boundaries is observed even at high applied strains. This indicates a lath boundary-favored plastic mechanism which survives up to high strain levels. The trends revealed in the strain distribution shown in Figure \ref{fig:map}d indeed corresponds to the strain distribution shown in \citep{morsdorf2016multiple}-Figure 2, obtained from experiments, where they have suggested that the there is no significant correlation between the strain localization zones and the corresponding Taylor factor (effect of crystallography).

\section{Conclusion} 

In this study, we used a habit-plane orientation-informed martensite model to examine the effect of the orientation-dependent yielding of lath martensite on the effective initial hardening behavior of martensitic steels. The results show that the initial plastic instability, i.e. 'micro-yielding', may be triggered by sliding of the soft inter-lath austenite films in favorably oriented packets, or related plastic mechanism that results in shearing at or near the substructure (lath, sub-block, block) boundaries. Therefore, the low stress level at which the boundary sliding occurs sets the macroscopic initial yield stress. The subsequent steep initial hardening is a reflection of the combined response of soft plastic packets, which undergo substructure boundary sliding, and harder ones, which are still elastic.
A gradual engagement of the less favorably oriented packets in plasticity results in a strongly heterogeneous fields in the microstructure inducing the observed extended elasto-plastic transition regime. The influence of changing the carbon content is accounted for by varying the lath strength and austenite volume fraction. It is shown that the magnitude of the work hardening rate increases with the strength of the laths, i.e. by increasing the strength contrast between the laths and austenite films. Similarly, the uniform elongation increases with the ultimate strength level of the microstructure, contrary to the classical trade-off between strength and ductility for advanced high strength steels.
 
In this work, we emphasized the significance of orientation-dependent yielding of martensite and its role in the high initial hardening rate of martensitic steels. We demonstrate that the effect may be considerably larger than was originally assumed. Thereby, we identified another mechanism possibly underlying the low yield point followed by very high initial hardening rate, next to the ones reported in literature, i.e. residual type II stresses, a high dislocation density resulting from the martensitic transformation, and (compositional, crystallographic, morphological) heterogeneities in the microstructure.  
Therefore, by taking the anisotropic yielding of lath martensite into account, more accurate predictions of the microstructural behavior can be achieved.

\section*{Acknowledgements}
This research was carried out under project number T17019b in the framework of the research program of the Materials Innovation Institute (M2i) (\href{www.m2i.nl}{www.m2i.nl}) supported by the Dutch government.

\newpage

\bibliography{/home/ador/Desktop/Papers/5thPaper/FinalVersion/library.bib}
\bibliographystyle{apalike}

\end{document}